\documentclass[a4paper,11pt]{article}
\usepackage{jinstpub} % for details on the use of the package, please see the JINST-author-manual
\usepackage{lineno}
\usepackage{siunitx}
\usepackage{caption}
\usepackage{subcaption,array}

\title{\boldmath A look-up table algorithm to model radiation damage effects in Monte Carlo events for HL-LHC experiments}

\author{M. Bomben and K. Nakkalil}
\affiliation{CNRS APC and Universit\'e Paris Cit\'e,\\
10 Rue Alice Domon et L\'eonie Duquet, 75013 Paris, France}
\emailAdd{marco.bomben@cern.ch}

\abstract{Radiation damage significantly impacts the performance of silicon tracking detectors in Large Hadron Collider (LHC) experiments such as ATLAS and CMS, with signal reduction being the most critical effect. Adjusting sensor bias voltage and detection thresholds can help mitigate these effects, but generating simulated data that accurately mirror the performance evolution with the accumulation of luminosity, hence fluence, is crucial.  The ATLAS collaboration has developed and implemented algorithms to correct simulated Monte Carlo (MC) events for radiation damage effects, achieving impressive agreement between collision data and simulated events.  
In preparation for the high-luminosity phase (HL-LHC), the demand for a faster ATLAS MC production algorithm becomes imperative due to escalating collision, events, tracks, and particle hit rates, imposing stringent constraints on available computing resources. This article outlines the philosophy behind the new algorithm, its implementation strategy, and the essential components involved.  The results from closure tests indicate that the events simulated using the new algorithm agree with fully simulated 
events at the level of few \%.  The first tests on computing performance show that the new algorithm is as fast as it is when no radiation 
damage corrections are applied.}


\arxivnumber{1234.56789} % Only if you have one

\begin{document}
\maketitle
\flushbottom

\section{Introduction}
\label{sec:intro}

Hybrid silicon pixel modules are at the core of the tracking detectors in High Energy Physics (HEP) experiments at the CERN Large Hadron Collider (LHC), see for example~\cite{IBL_paper,AtlasPixels}. The fast 
hadrons 
produced by LHC {\it pp} collisions can displace Si atoms and hence create deep states in the Si energy band gap. These deep states are responsible at macroscopic level of the increase of leakage current, the 
change in operational voltage and decrease of signal amplitude~\cite{MollTNS2018}. Operating the detector cold, adjusting sensor bias voltage and readout electronics thresholds and calibrations assure high 
hit efficiency despite the declining charge collection efficiency (CCE). It is also important to have simulated events mimicking  the evolution of CCE with the accumulation of luminosity, hence radiation damage 
fluence. The ATLAS collaboration developed and, since LHC Run~3, implemented algorithms~\cite{RadDamageDigi_2019} - the so called radiation damage digitizer - that starting from precise sensor electric field 
maps simulated using TCAD\footnote{Technology Computer Aided Design}  tools  can reproduce the evolution of cluster properties (charge, size) at \% level of agreement with data~\cite{EPS2023Bomben}. 

The LHC accelerator will be upgraded to a High Luminosity machine (HL-LHC)~\footnote{https://hilumilhc.web.cern.ch/content/hl-lhc-project} with instantaneous luminosity increasing by a factor of 3-5 with respect 
to LHC Run~3, with the goal to integrate over 10 years a dataset ten times larger than the one expected at the end of Run~3. The large increase in  instantaneous luminosity will translate into much larger 
events, particles and hits rates, in particular for the detector closest to the interaction point, the pixel detector; radiation damage doses and fluence will increase too, by almost a factor 10. For these reasons 
the ATLAS collaboration is preparing an upgraded pixel detector, the Inner Tracker (ITk) Pixel Detector~\cite{ATLASITkPixelTDR}, capable to assure the same if not even better performance of the original Pixel 
Detector but in a much harsher environment. 

The high luminosity phase of LHC will impose severe constraints also on computing resources. For example, despite the great performance of the radiation damage digitizer, it will have to be abandoned because 
it will be too demanding in terms of computing resources during the HL-LHC phase. In this paper a new lightweight algorithm for the radiation damage digitizer is presented. Its primary goal is to be 
as fast as possible, and to deliver precise predictions in terms of cluster charge and size as a function of the accumulated radiation damage. The algorithm is based on Look-Up tables~(section~\ref{sec:method}),  
which are prepared combining TCAD and MC simulations~(section~\ref{sec:creation}). The new algorithm has been tested performing several closure tests~(section~\ref{sec:closure}).  
Section~\ref{sec:conclusions} will end the article. 
%Also, watch out for the punctuation at the end of the equations.
%
%
%If you want some equations without the tag (number), please use the available
%starred-environments. For example:
%\begin{equation*}
%x = 1
%\end{equation*}

\section{Look-Up Table Method}
\label{sec:method}

In the original ATLAS radiation damage digitizer groupes of charge carriers produced by ionising particles are drifted towards the collecting electrodes, calculating their final position by integrating their movement 
based on electric field maps and mobility function; the signal amplitude on different pixel electrodes is then calculated using the Ramo theorem~\cite{ShockleyPot,Ramo}.
As said in the introduction this algorithm makes very precise predictions but it is significantly slower than the standard one (i.e. without radiation damage). 
In the new proposed algorithm the complicate dynamics of carriers will be replaced by scalar values, stored in so called Look-Up Tables (LUTs). For a charge $q$ deposited at depth $z$ inside the sensor bulk 
the final signal amplitude $q_{ind}$ and its final (``propagated'') position ($x_{prop},y_{prop},z_{prop}$) will be calculate using three LUTs, one for the $CCE$, one for the free path $\Delta z$ in $z$ and one for the 
tangent of Lorentz angle $\theta_{LA}(z)$; all LUTs will indeed depend on the deposition depth $z$. The signal amplitude and the propagated position are calculated as detailed in eq.~\ref{eq:dynamics}: 

%When possible, align equations on the equal sign. The package
%\texttt{amsmath} is already loaded. See \eqref{eq:x}.
\begin{equation}
\label{eq:dynamics}
\begin{aligned}
q_{ind} &= CCE(z) * q \,, \\
x_{prop} &= x + [\tan(\theta_{LA}(z)) \cdot \Delta z (z)] + \Delta^{diff} x\, \\
y_{prop} &= y + \Delta^{diff} y\, \\
z_{prop} & = z + \Delta z (z),
\end{aligned}
\end{equation}

where $\Delta^{diff} x,y$ is a Gaussian-distributed random number added to simulate the effect $x,y$ of diffusion. More details on the method 
can be found in ref.~\cite{s24123976}.

\section{LUT Creation}
\label{sec:creation}

In order to create the three LUTs it is necessary to have a precise simulation of the signals created by ionising particles in pixel detectors. In this work the task was accomplished using  the Allpix$^2$ 
MC simulation framework~\cite{Allpix2}. Within Allpix$^2$ the response to radiation of silicon detectors  can be simulated  with great detail, thanks to the possibility of controlling in a modular way 
all involved processes, from charge deposition to signal digitization and construction, including charge carrier propagation. 

\begin{figure}
  \centering
   \includegraphics[width=0.45\textwidth]{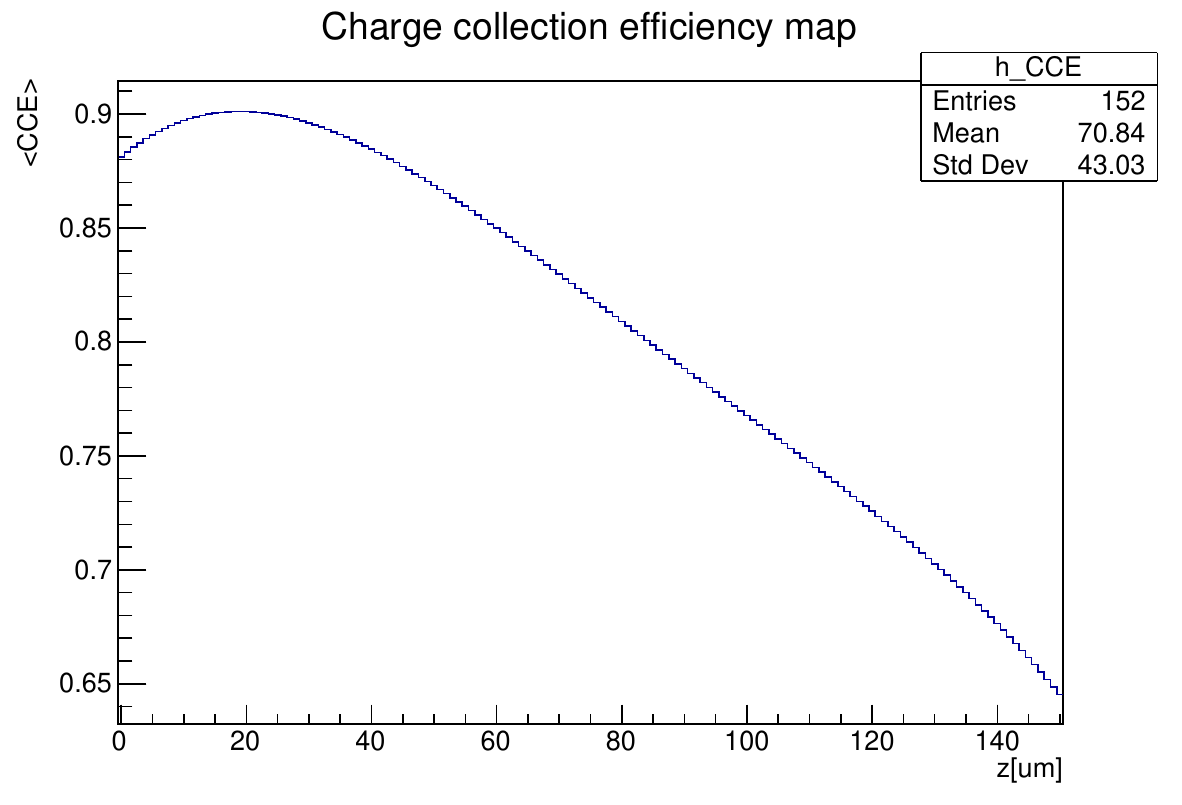}
    \includegraphics[width=0.45\textwidth]{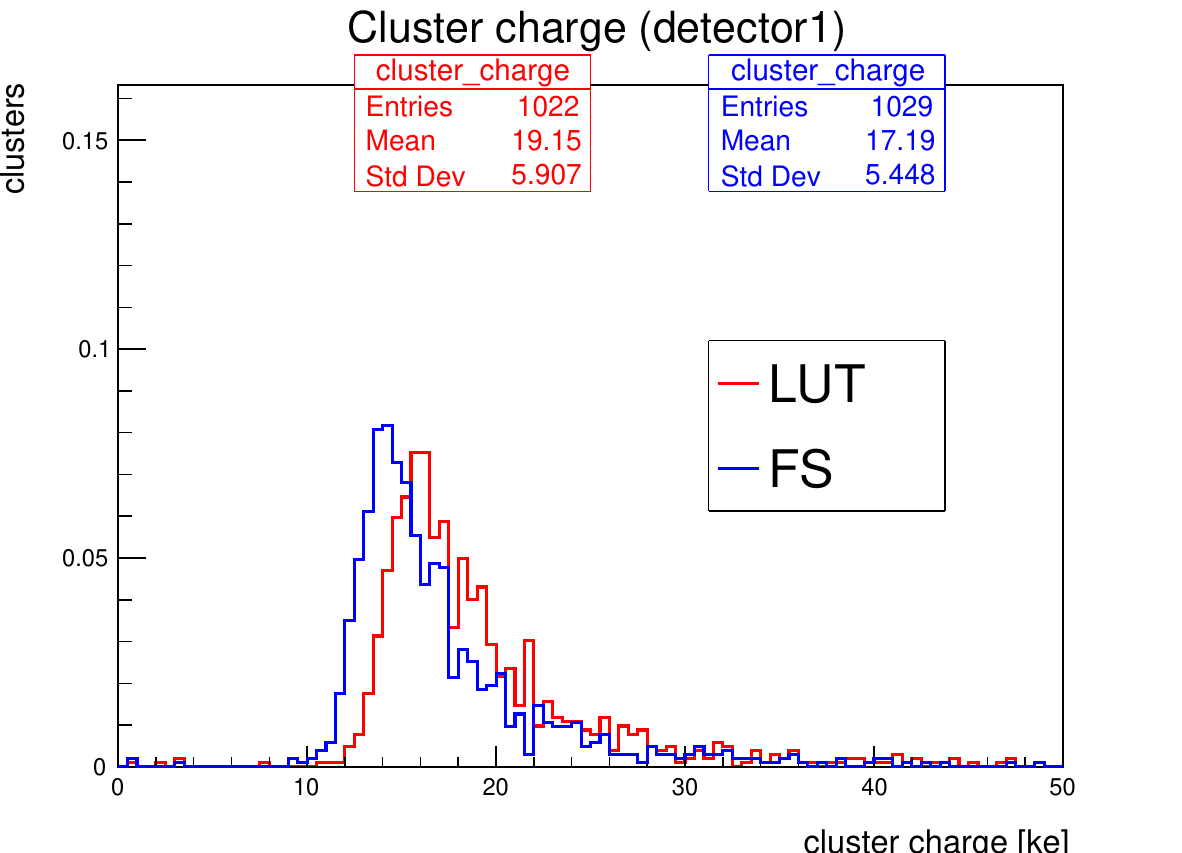}
     \includegraphics[width=0.45\textwidth]{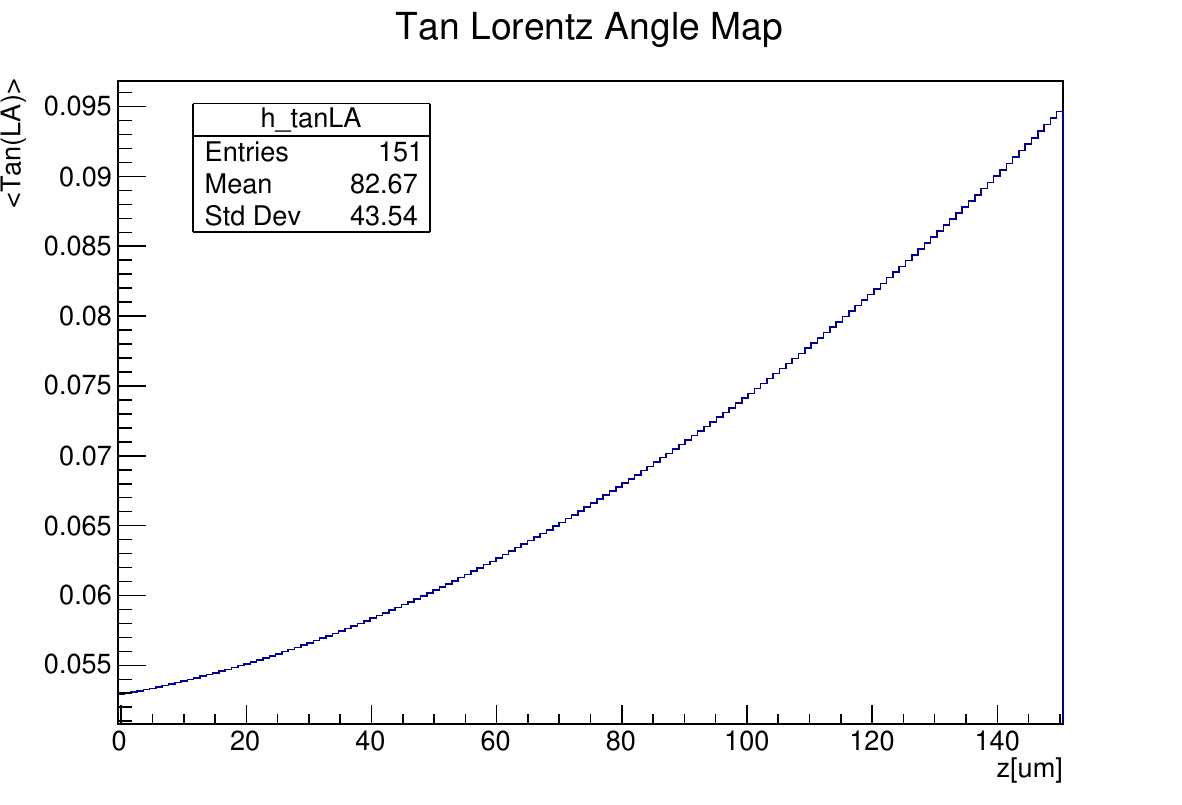}
     \includegraphics[width=0.45\textwidth]{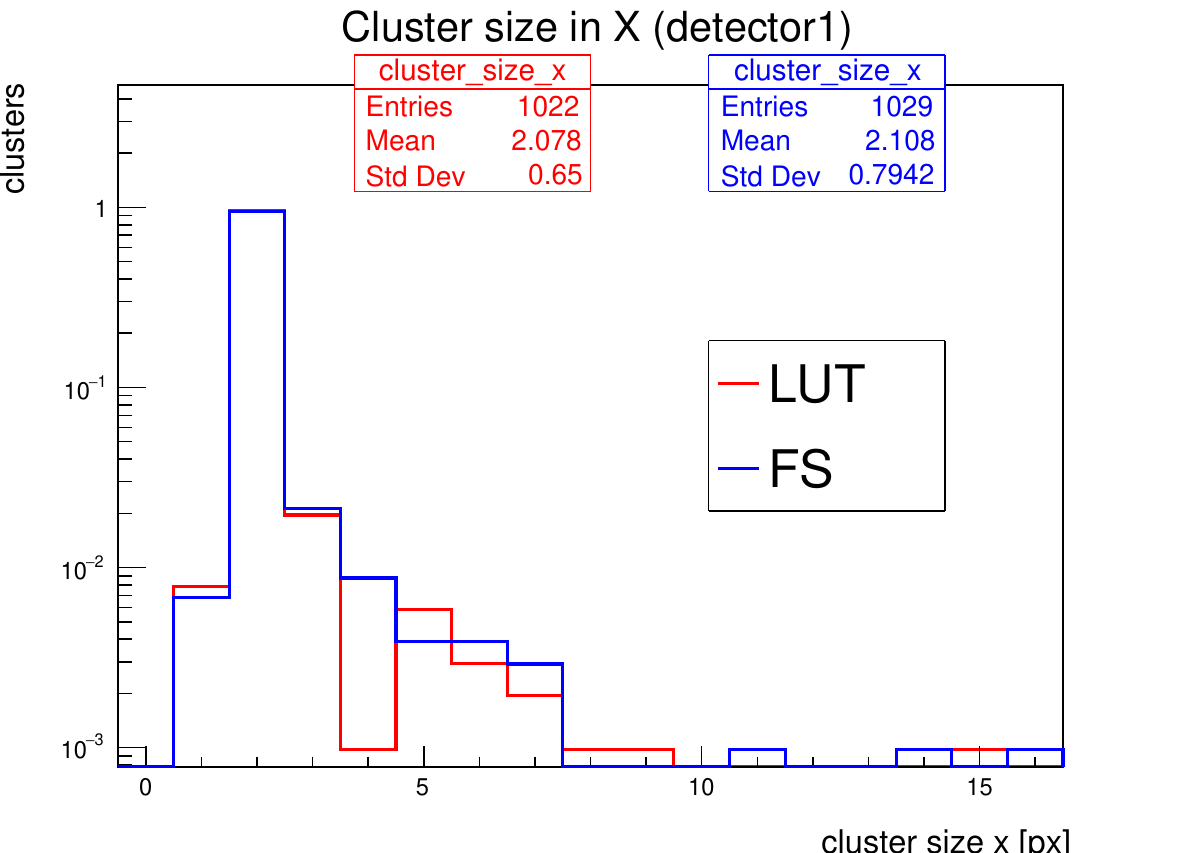}
        \includegraphics[width=0.45\textwidth]{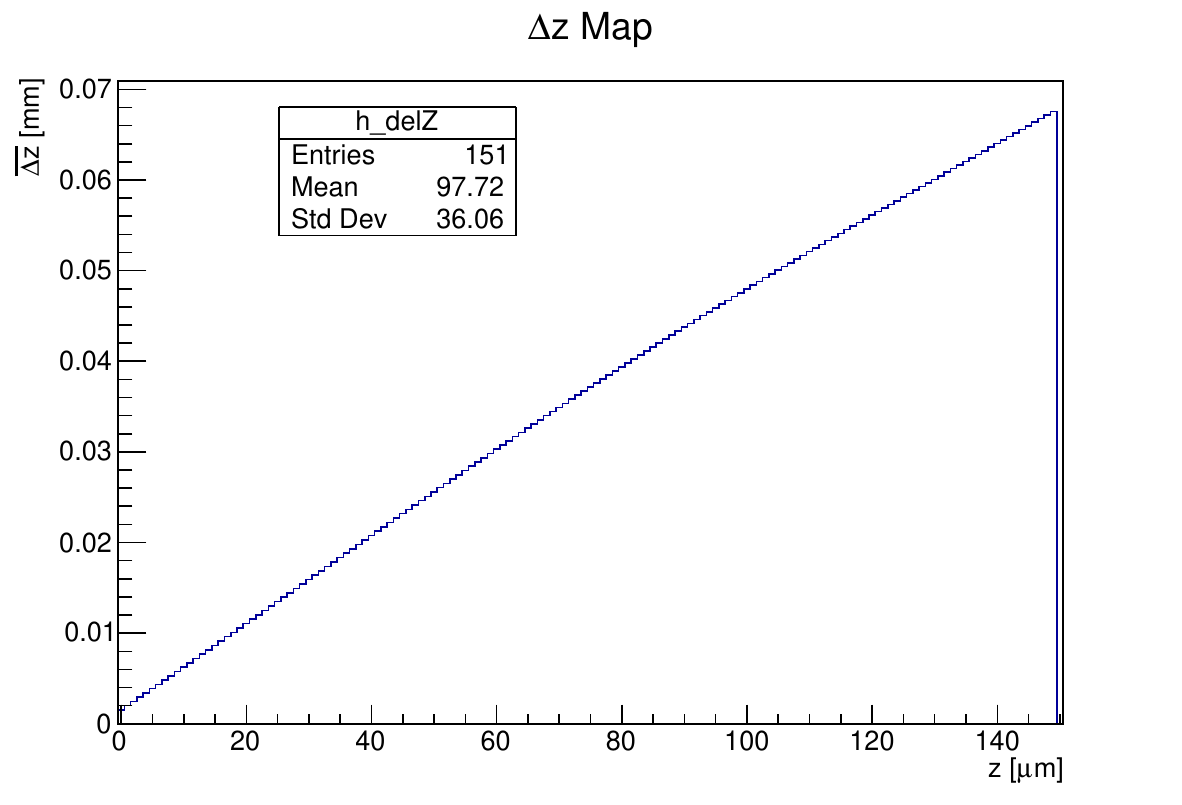}
       \includegraphics[width=0.45\textwidth]{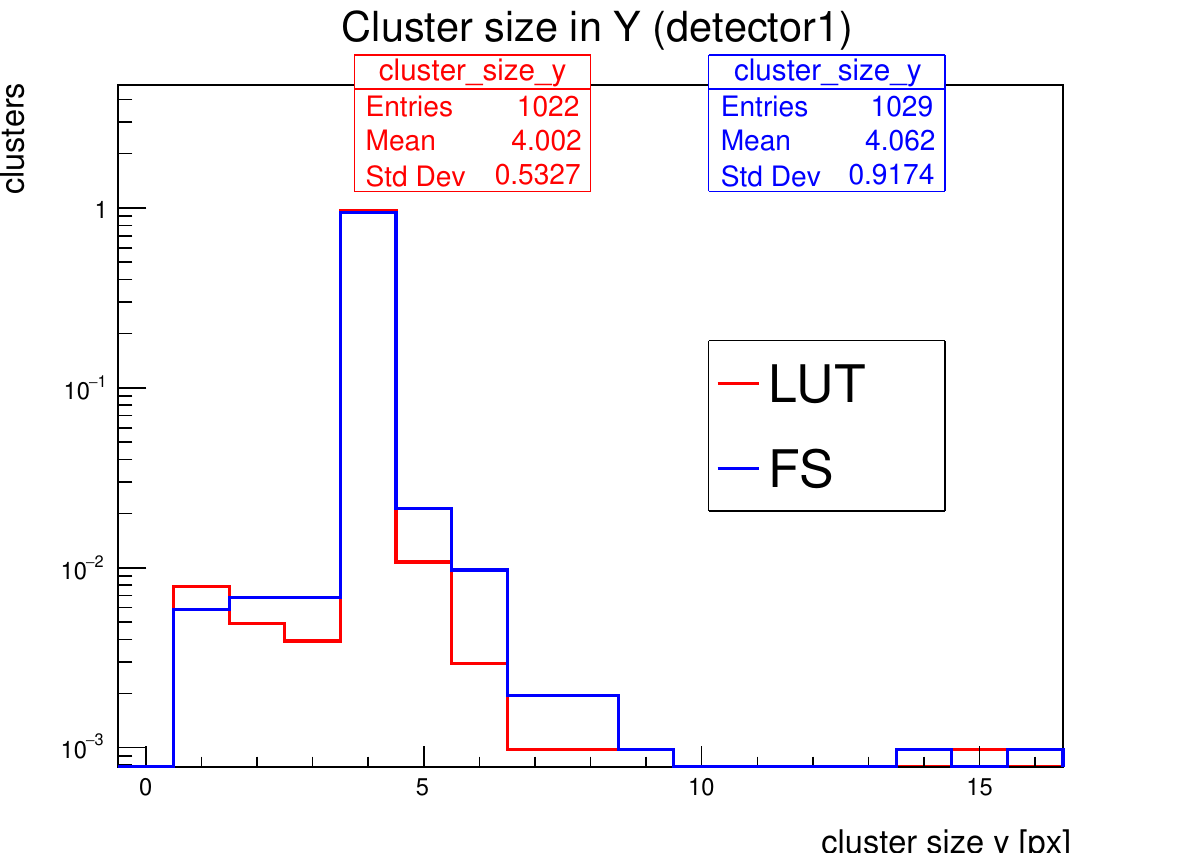}
\caption{ \label{fig:mixed}(Left) LUTs for irradiated planar pixels, (\SI{150}{\micro\meter} thick, 50$\times$50~\SI{150}{\micro\meter}$^2$, $\Phi$ = $1 \times 10^{15}$~\SI{}{n_{eq}/cm^2}, V$_{\rm bias}$ = 400~V).
 (top) CCE; (mid) tangent    of Lorentz ange; (bottom) average free path. \\(Right) Comparison of LUT and FS events at $\eta$=1. (top) cluster charge; (mid) transverse cluster size; (bottom) longitudinal cluster size.}
\end{figure}

Using   Allpix$^2$  it was possible to simulate events for a \SI{150}{\micro\meter} thick n-on-p planar pixel sensors with a pitch of 50$\times$50~\SI{150}{\micro\meter}$^2$. The electric field profile map was 
taken from a TCAD simulation that included radiation damage effects corresponding to a fluence $\Phi$ of $1 \times 10^{15}$~\SI{}{n_{eq}/cm^2}; the TCAD model for radiation 
damage~\cite{FOLKESTAD201794} developed by the LHCb collaboration was used. The Ramo potential map was calculated too using TCAD tools.
The resulting LUTs are reported in Figure~\ref{fig:mixed} for a bias voltage of V$_{\rm bias}$ = 400~V.

%\begin{figure}[htbp]
  % \centering
 %  \includegraphics[width=0.3\textwidth]{figures/CCE.pdf} % requires the graphicx package
    %  \includegraphics[width=0.3\textwidth]{figures/LA.pdf} % requires the graphicx package
       %  \includegraphics[width=0.3\textwidth]{figures/dz.pdf} % requires the graphicx package
  % \caption{LUTs for irradiated pixels (\SI{150}{\micro\meter} thick, 50$\times$50~\SI{150}{\micro\meter}$^2$, $\Phi$ = $1 \times 10^{15}$~\SI{}{n_{eq}/cm^2}, V$_{\rm bias}$ = 400~V). (left) CCE; (mid) tangent 
  % of Lorentz ange; (right) average free path. }
   %\label{fig:luts}
%\end{figure}  

%\begin{figure}
  %  \centering
  %  \begin{tabular}{cc}
  %  \begin{subfigure}{0.33\textwidth}
      %  \centering
    %    \includegraphics[height=1.7in]{figures/CCE.pdf}
      %  \caption{}
   % \end{subfigure}%
   % &
   %     \begin{subfigure}{0.33\textwidth}
     %   \centering
     %   \includegraphics[height=1.7in]{figures/LA.pdf}
     %   \caption{}
  %  \end{subfigure}%
  %  \\
    % uncomment next line
    % \multicolumn{x}{c}{%
  %  \begin{subfigure}{0.5\textwidth}
      %  \centering
     %   \includegraphics[height=1.7in]{figures/dz.pdf}
   %     \caption{}
    %\end{subfigure}%}// <- uncomment
   % % comment next line
    %& \\
    %\end{tabular}
      % \caption{LUTs for irradiated pixels (\SI{150}{\micro\meter} thick, 50$\times$50~\SI{150}{\micro\meter}$^2$, $\Phi$ = $1 \times 10^{15}$~\SI{}{n_{eq}/cm^2}, V$_{\rm bias}$ = 400~V). (a) CCE; (b) tangent 
   %of Lorentz ange; (c) average free path. }
   %\label{fig:luts}
%\end{figure}

From the CCE LUT is visible the effect of signal screening of holes in the region between pixel side (\SI{0}{\micro\meter}) and \SI{20}{\micro\meter} away from it; at \SI{80}{\micro\meter} from the pixel 
the free path is reduced to half; the largest deflection is of course expected from carriers produced at the backside.

\section{Closure Tests}
\label{sec:closure}

Lacking experimental data from irradiated pixel modules, a series of closures tests were conducted in order to validate the LUT method and ingredients. In particular, a comparison of events simulated using the 
LUT method with Fully Simulated (FS) ones was performed. The distribution of cluster charge and cluster size in both projections - transverse and parallel to the magnetic field - 
 were compared between LUT and FS events; more details in~\cite{s24123976}.
Here the results of a closure test for the device and conditions mentioned in sect.~\ref{sec:creation} are reported. The module was tilted  at different angles with respect to the simulated $\pi^+$ beam to 
emulate the data-taking conditions in ATLAS. In figure~\ref{fig:mixed} a comparison for pseudorapidity\footnote{$\eta = - \ln (\tan (\theta / 2 ))$, where $\theta$ is the polar angle with 
respect to the direction of the beam} $\eta = 1$.

%\begin{figure}[htbp]
  % \centering
   %\includegraphics[width=0.3\textwidth]{figures/cluster_charge_LUT_vs_FS.pdf} 
   %\includegraphics[width=0.3\textwidth]{figures/cluster_size_x_LUT_vs_FS.pdf}
     %\includegraphics[width=0.3\textwidth]{figures/cluster_size_y_LUT_vs_FS.pdf}
   %\caption{ \label{fig:closuretest}Comparison of LUT and FS events at $\eta$=1. (left) cluster charge; (center) transverse cluster size; (right) longitudinal cluster size.}
%\end{figure} 

%\begin{figure}
  %  \centering
    %\begin{tabular}{cc}
   % \begin{subfigure}{0.33\textwidth}
      %  \centering
        %\includegraphics[height=1.7in]{figures/cluster_charge_LUT_vs_FS.pdf}
        %\caption{}
   % \end{subfigure}%
    %&
       % \begin{subfigure}{0.33\textwidth}
        %\centering
        %\includegraphics[height=1.7in]{figures/cluster_charge_LUT_vs_FS.pdf}
        %\caption{}
   % \end{subfigure}%
    %\\
    % uncomment next line
    % \multicolumn{x}{c}{%
    %\begin{subfigure}{0.5\textwidth}
      %  \centering
       % \includegraphics[height=1.7in]{figures/cluster_charge_LUT_vs_FS.pdf}
       % \caption{}
   % %\end{subfigure}%}// <- uncomment
    % comment next line
   % & \\
    %\end{tabular}
    %\caption{\label{fig:closuretest}Comparison of LUT and FS events at $\eta$=1. (a) cluster charge; (b) transverse cluster size; (c) longitudinal cluster size.}
  %\end{figure}

The level of agreement is remarkable for both projections of cluster size. LUT based events are characterised by an average cluster charge 10\% larger than FS ones; this will be investigated. 
Overall the agreement can be considered satisfactory.

\begin{figure}[htbp]
   \centering
   \includegraphics[width=0.5\textwidth]{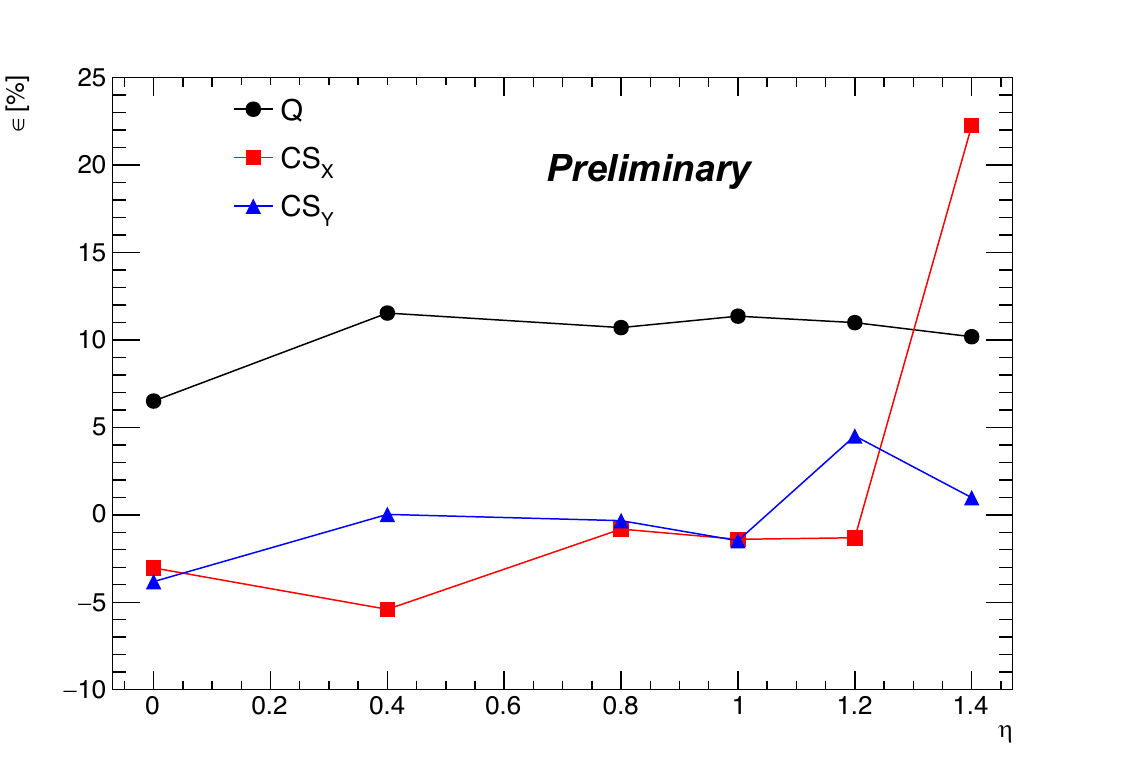} 
   \caption{ \label{fig:summaryclosure}Relative percentage difference $\epsilon$ between LUT and FS events at different $\eta$ values. For all observables $\epsilon$ is calculated as the 
   difference between mean values of the observable distribution of LUT and FS events divided by the value of FS events.}
\end{figure}

The study has been repeated for several $\eta$ values. A relative difference $\epsilon$ as been calculated as:

\begin{equation}
\label{eq:epsilon}
\epsilon = \dfrac{\left<O\right>_{LUT}-\left<O\right>_{FS}}{\left<O\right>_{FS}}
\end{equation}

where $\left<O\right>$ is the average value of the distribution of observable $O$ which is either cluster cluster charge $Q$ or transverse/longitudinal cluster size $CS_{X,Y}$. 
Results are reported in figure~\ref{fig:summaryclosure}.

The agreement in both projections of cluster size is at a few \% level which is extremely promising; cluster charge is somewhat larger in LUT than FS events. At $\eta = 1.4$ the agreement for transverse 
cluster size is less optimal: this is similar to what reported in~\cite{s24123976} for a different simulated sensor. The situation is being investigated. 

\section{Conclusions and Outlook}
\label{sec:conclusions}

Radiation damage is the limiting effect of pixel detector performance at hadron colliders. In this paper a novel algorithm to include these effects in MC simulated events has been discussed. 
Closures tests indicate it is quite precise in reproducing cluster properties like charge and size. In a  preliminary test on computing performance the production of simulated events using the LUT method 
was as fast as the basic version, i.e. without radiation damage. This was expected since the algorithmic  complexity of the LUT algorithm is basically the same of the original one.

\acknowledgments

The authors want to thank Jeff Dandoy from Carleton University for useful discussions and for helping us in optimising the algorithm.

% Bibliography

%% [A] Recommended: using JHEP.bst file
 \bibliographystyle{JHEP}
 \bibliography{biblio.bib}

%% or
%% [B] Manual formatting (see below)
%% (i) We suggest to always provide author, title and journal data or doi:
%% in short all the informations that clearly identify a document.
%% (ii) please avoid comments such as "For a review'', "For some examples",
%% "and references therein" or move them in the text. In general, please leave only references in the bibliography and move all
%% accessory text in footnotes.
%% (iii) Also, please have only one work for each \bibitem.

%\begin{thebibliography}{99}

%\bibitem{a}
%Author,
%\emph{Title},
%\emph{J. Abbrev.} {\bf vol} (year) pg.

%\bibitem{b}
%Author,
%\emph{Title},
%arxiv:1234.5678.

%\bibitem{c}
%Author,
%\emph{Title},
%Publisher (year).

%\end{thebibliography}
\end{document}